\documentclass[11pt,dvips]{article}

\usepackage{epsfig,times}
\usepackage{picinpar}
\usepackage{wrapfig}
\usepackage{floatflt}
\makeatletter
\renewcommand{\section}{\@startsection{section}{1}{0in}
	{0.4\baselineskip}{0.1\baselineskip}{\Large\bf}}
\renewcommand{\subsection}{\@startsection{subsection}{2}{0in}
	{0.25\baselineskip}{-\baselineskip}{\large\bf}}
\renewcommand{\subsubsection}{\@startsection{subsubsection}{3}{0in}
	{0.1\baselineskip}{-\baselineskip}{\normalsize\bf}}
\makeatother
\newcommand{\apj}{ApJ}
\newcommand{\apjl}{ApJ}

\newcommand{\nat}{Nature}

\begin{document}

%
\thispagestyle{myheadings}
%
\markright{OG 2.1.13}
\begin{center}
%
{\LARGE \bf VHE Gamma Rays from PKS 2155--304}
\end{center}

\begin{center}
%
%
{\bf P.M.~Chadwick, K.~Lyons, T.J.L.~McComb, K.J.~Orford,
J.L.~Osborne, S.M.~Rayner, S.E.~Shaw, and K.E.~Turver}\\
{\it Department of Physics, Rochester Building, Science Laboratories,
University of Durham, Durham, DH1~3LE, U.K.}
\end{center}

\begin{center}
{\large \bf Abstract\\}
\end{center}
\vspace{-0.5ex}
%
%

The X-ray selected BL Lac PKS 2155--304 has been observed using the
University of Durham Mark 6 very high energy gamma ray telescope during
1998. We find no evidence for TeV emission during these recent
observations when the X-ray flux was observed to be low. We have
reconsidered our measurements made in 1997 November when PKS 2155--304
was in a bright X-ray state and extended X-ray and GeV gamma ray
observations were made as part of a multiwavelength campaign.
Comparisons are made of the VHE emission during this time with the
available data from other wavelengths.

\vspace{1ex}

%
%

\section{Introduction}

Evidence exists for at least four close X-ray selected BL Lacs as
sources of episodic TeV gamma ray emission (Mrk 421 -- Punch et al.
1992, Mrk 501 -- Quinn et al. 1996, 1ES 2344+514 -- Catanese et al. 1997
and PKS 2155--304 -- Chadwick et al. 1999). The discovery of PKS
2155--304 as a VHE gamma ray source was made with the University of
Durham Mark 6 telescope on the basis of observations lasting 40 hrs in
1996 and 1997. These results suggested a time variable emitter with the
strongest emission in 1997 November at the time of a successful
multiwavelength campaign (Chadwick et al. 1999). 
\begin{wraptable}[23]{l}{8cm}
\begin{center}
\begin{tabular}{@{}lc}
\hline \hline
Date & No. of \\
& scans \\
& ON source \\ \hline
1998 July 22 & 1 \\
1998 August 18 & 5 \\
1998 August 19 & 7 \\
1998 August 20 & 4 \\
1998 September 15 & 5\\
1998 September 16 & 2\\
1998 September 17 & 4 \\
1998 September 19 & 3 \\
1998 October 11 & 2 \\
1998 October 13 & 2\\
1998 October 16 & 1 \\ \hline
\end{tabular}
\end{center}
\caption{Observing log for our observations of PKS 2155--304 during
1998.} \label{observing_log} 
\end{wraptable}
We have extended the observations of PKS 2155 with further measurements
in 1998 August -- November. 

The details of the measurements at X-ray energies in 1997 November are
now available from both {\it RXTE} (Vestrand \& Sreekumar 1999) and {\it
BeppoSAX} (Chiappetti et al. 1999), together with data from {\it
CGRO}/EGRET (Vestrand \& Sreekumar 1999). During the 36 hour observation
with {\it BeppoSAX} a short interval (2 hrs) of simultaneous X-ray
and TeV observations occurred.

We here report the results of our 1998 measurements and reconsider our
1997 November data in the light of the recently available X-ray results
from the multiwavelength campaign. All measurements reported here have
been made with the University of Durham Mark 6 gamma ray telescope
operating at Narrabri, NSW, Australia. The telescope has been described
in detail by Chadwick et al. (1997) and Armstrong et al. (1999).

\section{New measurements in 1998}

Observations in 1998 have involved 9 hrs of exposure ON source and an
equal amount OFF source during 1998 August, September and October and
the observing log is summarized in Table \ref{observing_log}. All
analyses reported here have been completed using the same procedures and
background suppression used in our previous work on PKS 2155--304
described by Chadwick et al. (1999).

\begin{table*}[t]
\begin{center}
\begin{tabular}{@{}lrrrr}
\hline \hline
& On & Off & Difference & Significance \\ \hline
Number of raw events & 171723 & 173203 & $-1480$ & $-2.5~\sigma$ \\
Number of size and & 97279 & 97493 & $-214$ & $-0.48~\sigma$ \\
distance selected events & & & & \\
Number of shape & 6260 & 6053 & 207 & $1.9~\sigma$ \\
selected events & & & & \\
Number of shape and & 950 & 992 & $-42$ & $-0.10~\sigma$ \\
{\it ALPHA} selected events & & & & \\ \hline
\end{tabular}
\end{center}
\caption{The results of various event selections for the 1998 PKS 2155--304
data.}
\label{result_table}
\end{table*}
\begin{wrapfigure}[19]{l}{9cm}
\centerline{\psfig{file=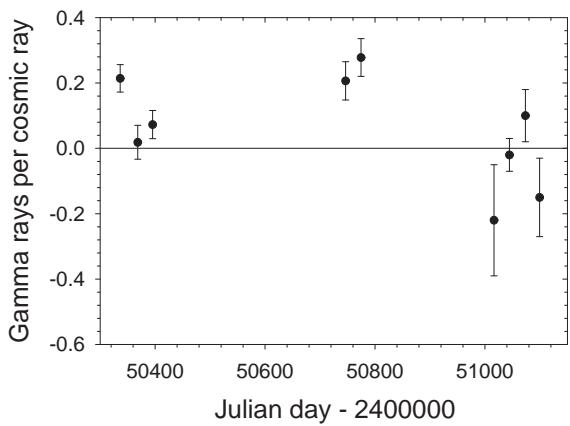,height=7cm}} 
\caption{The measured VHE
gamma ray flux above 300 GeV from PKS 2155--304 averaged over observing
periods of approximately 10 days.} \label{timehistory} 
\end{wrapfigure}
We find no evidence for emission of TeV gamma rays throughout the
observation --- see Table \ref{result_table} for the summary of the data
analysis. We calculate a time-averaged $3 \sigma$ flux limit of $4.0
\times 10^{-11}~{\rm cm}^{-2}~{\rm s}^{-1}$ at an energy theshold of 300
GeV for 1998 August -- October.

We show in Figure \ref{timehistory} the time averaged VHE gamma ray
fluxes (normalized to the cosmic ray counting rate --- see Chadwick et
al. 1999) for all dark periods during which we have observed PKS
2155--304. We note that according to the X-ray measurements made with
the ASM on board {\it RXTE}\footnote{Available on the web at
\mbox{http://space.mit.edu/XTE/asmlc/srcs/pks2155-304.html}.} PKS 2155--304
was in an X-ray low state during our measurements in 1998 August --
October. Our failure to detect VHE emission during 1998 is thus
consistent with the hypothesis that the X-ray and VHE gamma-ray emission
from PKS 2155--304 are correlated (Chadwick et al. 1999). Similar
behaviour is observed in the VHE and X-ray emission from Mrk 421 and Mrk
501.

\section{Multiwavelength Observations in 1997 November}

\begin{wrapfigure}[25]{l}{9cm}
\centerline{\psfig{file=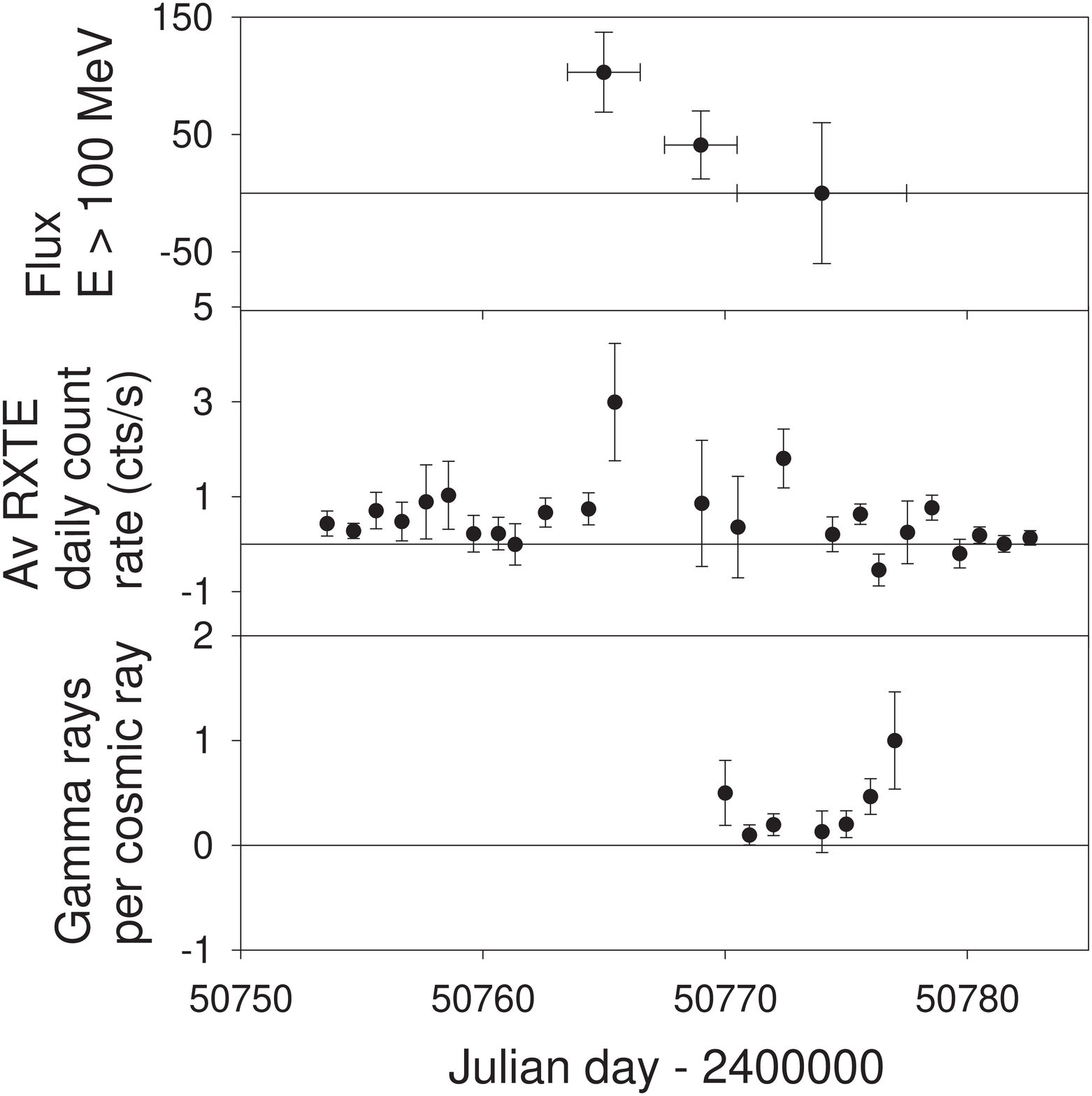,height=8cm}}
\caption{The GeV gamma rays recorded with EGRET (Vestrand and Sreekumar
1999 -- upper panel), X-rays recorded with ASM on {\it RXTE} (centre
panel) and VHE gamma ray (present work -- bottom panel) during the 1997
November observations of PKS 2155--304.} \label{xgamma} 
\end{wrapfigure}

TeV gamma ray data are available from observations
between 17--25 November (see Chadwick et al. 1999). These observations
were made as part of a multiwavelength campaign
involving GeV gamma ray and X-ray measurements. The GeV gamma ray
measurements were made using {\it CGRO}/EGRET from 1997 Nov 10 -- Nov 23
and showed strong GeV emission during the first half of the viewing
period (Vestrand \& Sreekumar 1999). X-ray observations using PCA and
HEXTE on board {\it RXTE} were made during 1997 Nov 20 -- 22 (Vestrand
\& Sreekumar 1999) and ASM observations are available throughout. {\it
BeppoSAX} observed this object for about 1.5 days during 1997 Nov 22 --
24 (Chiappetti et al. 1999). The X-ray and gamma ray observations
clearly show that PKS 2155--304 was in an active flaring state during
the middle of November 1997, with X-ray and gamma ray fluxes being as
high as ever previously detected --- see Figure \ref{xgamma}. Recently
the CANGAROO group have published data from PKS 2155--304 taken between
1997 Nov 24 and 1997 Dec 1 (Roberts et al. 1999). They fail to detect
any TeV emission from PKS 2155--304 during this period, quoting a flux
limit $< 9.5 \times 10^{-12}~{\rm cm}^{-2}~{\rm s}^{-1}$ at an energy
threshold of 1.5 TeV. Given the non-overlapping observation periods,
the possibility of time variation in the emission and the different
thresholds of the telescopes, this null result is not in conflict with
our detection.
 
Our TeV observations, averaged over the total dataset for 1997 November,
indicated the strongest emission during any of our observations of PKS
2155 to date (Chadwick et al. 1999). The time averaged flux of VHE gamma
rays from PKS 2155-304 for our observations in 1997 November is $(6.0
\pm 2.0_{stat} \pm 3.0_{sys}) \times 10^{-11}~{\rm cm}^{-2}~{\rm
s}^{-1}$ for an energy threshold of 300 GeV. The GeV and X-ray
observations suggest that a large outburst occurred in early November
prior to the TeV observations (Vestrand \& Sreekumar 1999). The TeV
observations do not contradict this idea.

\begin{figure}[th]
\centerline{\psfig{file=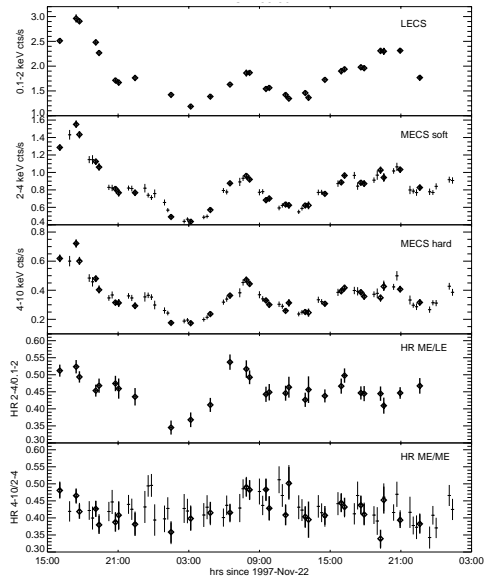,height=8cm}} 
\centerline{\psfig{file=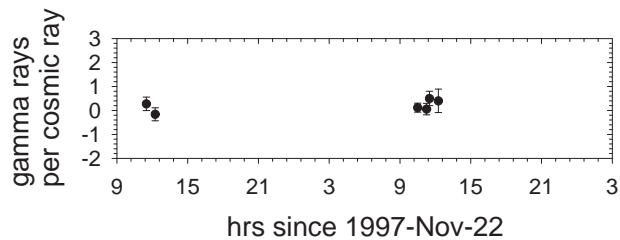,height=4.4cm}}
\caption{(a) The X-ray light curves (upper three panels) and hardness ratios
(lower two panels) recorded with {\em BeppoSAX} during the 1997 November
observations of PKS 2155--304 (taken from Chiappetti et al. 1999). Also
shown (b) are the VHE gamma ray results from the present work.}
\label{bepposax}

\end{figure}

We have considered on a day by day basis the information available from
ground based Cerenkov telescopes, {\it CGRO}/EGRET, {\it RXTE} and {\it
BeppoSAX}. The only truly contemporaneous data were recorded by {\it
BeppoSAX} and the Mark 6 telescope on 1997 November 23 between 1100 and
1300 hrs UTC. We reproduce the X-ray data from the paper of Chiappetti
et al. (1999) and focus on the interval with VHE observations --- see
Figure \ref{bepposax}(a). 

Our VHE observations occur at the time which Chiappetti et al. (1999)
define as a region of low X-ray intensity defined on the basis of the
MECS (medium energy) count rate, beginning about 2 hours after the peak
of the second X-ray flare detected by {\it BeppoSAX}. We show in Figure
\ref{bepposax}(b) the results of our VHE observations for individual 15
min scans on 1997 Nov 23, along with data taken on 1997 Nov 22 which
were obtained about three hours before the {\it BeppoSAX} observation
commenced. We have no evidence for strong flaring activity within the
simultaneous VHE data obtained on 1997 Nov 23, consistent with the low
X-ray state. The VHE data taken on 1997 Nov 22 are at the same activity
level as on Nov 23. The X-ray data show that an X-ray flare peaked about
2 hrs after our VHE observation on November 22nd finished and that the
typical time scale for X-ray flaring is such that the flare is likely to
have commenced after our observation terminated. The VHE data yield a
flux of $(2.0 \pm 5.0_{stat} \pm 1.0_{sys}) \times 10^{-11}~{\rm
cm}^{-2}~{\rm s}^{-1}$ for the observation on 1997 Nov 22 and $(7.0 \pm
4.5_{stat} \pm 3.5_{sys}) \times 10^{-11}~{\rm cm}^{-2}~{\rm s}^{-1}$
for the observation on 1997 Nov 23, both at an estimated energy
threshold of 300 GeV.

\section{Discussion}

In common with the VHE observations of the Northern hemisphere blazars,
the VHE emission from PKS 2155--304 is time variable and appears to be
correlated with X-ray emission, consistent with the unified model of
AGNs. We will continue to monitor this object with the Mark 6 telescope
and we plan to observe it intensively when there is any indication of
flaring activity at X-ray energies from the ASM quick look analysis or
at any other wavelengths. The 1997 multiwavelength campaign has yielded
the first simultaneous observations of VHE gamma rays from this object
at the same time as data from high throughput X-ray instruments.
Unfortunately, the VHE coverage did not overlap one of the short (2hr
duration) X-ray flares. However, the sensitivity of the Mark 6 telescope
is such that a flare of comparable X-ray intensity to that of 1997 Nov
22 should be resolvable in its VHE emission.

We are grateful to the UK Particle Physics and Astronomy Research
Council for support of the project. This paper uses quick look results
provided by the ASM/{\it RXTE} team.

\vspace{1ex}
\begin{center}
{\Large\bf References}
\end{center}
%
Armstrong, P., et al. 1999, Exp. Astron., in press\\
Catanese, M., et al. 1997, \apjl, 487, L143\\
Chadwick, P. M., et al. 1997, in Towards a Major Atmospheric
Cerenkov Detector - V, ed. O. C. de Jager (Potchefstroom: Potchefstroom
University for CHE), p. 172\\
Chadwick, P. M., et al. 1999, \apj, 513, 161\\
Chiappetti, L., et al. 1999, \apj, in press\\
Punch, M., et al. 1992, \nat, 358, 477\\
Quinn, J., et al. 1996, \apjl, 456, L83\\
Roberts, M. D., et al. 1999, A\&A, 343, 691\\
Vestrand, W. T. \& Sreekumar, P. 1999, Astropart. Phys., in press\\

\end{document}